\documentclass{PoS}
\newcommand{\bvec}[1]{\boldsymbol{#1}}
\newcommand{\Expect}[1]{\left\langle{#1}\right\rangle}
\newcommand{\no}{\nonumber}

\usepackage{amsmath,amssymb,amscd,mathrsfs,fancybox,amsthm}
\usepackage{graphicx}
\usepackage{dcolumn}
\usepackage{bm}
\usepackage{bbold}
\title{Direct Proof of Reflection Positivity of Free Overlap Dirac Fermion}

\ShortTitle{Direct Proof of Reflection Positivity of Free Overlap Dirac Fermion}

\author{Yoshio Kikukawa\\
	Institute of Physics, the University of Tokyo, Tokyo 153-8902, Japan\\
	E-mail: \email{kikukawa@hep1.c.u-tokyo.ac.jp}
}%

\author{\speaker{Kouta Usui}\\
        Department of Physics, the University of Tokyo 113-0033, Japan\\
        Institute for the Physics and Mathematics of the Universe (IPMU),   the University of Tokyo, Chiba 277-8568, Japan\\
        E-mail: \email{kouta@hep-th.phys.s.u-tokyo.ac.jp}}

\abstract{It is shown that free lattice Dirac fermions defined by overlap Dirac operator  fulfill the  Osterwalder-Schrader reflection positivity condition with respect to the link-reflection.  }

\FullConference{The XXVIII International Symposium on Lattice Field Theory, Lattice2010\\
		June 14-19, 2010\\
		Villasimius, Italy}

\begin{document}

\section{Introduction}
In this short article,  we will examine the reflection positivity \cite{Osterwalder:1973dx,Osterwalder:1974tc,Osterwalder:1977pc} 
of lattice fermions defined through 
overlap Dirac operator \cite{Neuberger:1997fp, Neuberger:1998wv}, 
a gauge-covariant solution to the GW relation \cite{Neuberger:1997fp, Neuberger:1998wv,Ginsparg:1981bj, Hasenfratz:1998ri, Hasenfratz:1998jp, Luscher:1998pqa}, which has been derived in the five-dimensional 
domain wall approach \cite{Kaplan:1992bt, Shamir:1993zy, Furman:1994ky}.
The reflection positivity of the GW fermions is not fully understood yet \cite{Luscher:2000hn, Creutz:2004ir, Mandula:2009yd},
while that of Wilson fermions has been rigorously proved in various ways including the gauge interacting case
\cite{Osterwalder:1977pc, Luscher:1976ms, Menotti:1987cq}.
In the following, It will be shown rigorously that free overlap Dirac fermion fulfills the reflection positivity with respect to the link-reflection.  
In ref \cite{Luscher:2000hn}, L\"{u}scher discussed the unitarity property of free overlap Dirac fermion
by investigating the positivity through the spectral representation of free propagator and concluded that 
free overlap Dirac fermion has a good unitarity property. 
Our direct proof of the reflection positivity given here is consistent with this observation.
Our proof 
will be also 
extended to the non-gauge models with interactions such as 
chiral Yukawa models.  
For gauge models, however, a proof of reflection positivity, if any, 
seems to be more involved and we will leave it for future study. 

\vspace{.5em}
\noindent
\section{Reflection positivity} 
Reflection positivity  
is a sufficient condition for 
reconstructing a quantum theory in the canonical formalism, i.e. 
the Hilbert space of state vectors and 
the Hermitian Hamiltonian operator acting on the state vectors, 
from the lattice model defined in the Euclidean space
Let us formulate the reflection positivity condition for lattice Dirac fermions.

%
We assume a finite lattice
$\Lambda=[-L+1,L]^4\subset \mathbb{Z}^4$ in the lattice unit $a=1$, and 
impose \textit{anti}-periodic boundary condition in the time direction, 
and periodic boundary conditions in the space directions. 
The fermionic action is 
defined in the bilinear form
\begin{align}
\label{eq:Dirac-fermion-action}
A(\bar\psi,\psi)=\sum_{x\in\Lambda}\bar\psi(x)D_L\psi(x),
\end{align}
with  a lattice Dirac operator $D_L$\footnote{
The reader might prefer the sign convention where $A=-\bar\psi D_L \psi$ 
in stead of \eqref{eq:Dirac-fermion-action}.
These two sign convention in the fermionic action are connected to each other by the transformation
$\bar\psi'=i\bar\psi$ and $\psi'=i\psi$.
We have chosen this sign convention for the sake of the proof of the reflection positivity 
following \cite{Osterwalder:1977pc}. As we will see 
in section 3, 
this sign convention is suitable to prove the statement (iv) $\Delta A\in\mathcal{\bar P}$. 
}.
The kernel of the Dirac operator should be written as 
\begin{align}
D_L(x,y)=\sum_{n\in\mathbb{Z}^4}(-1)^{n_0}D(x+2nL,y),\quad x,y\in\Lambda,\label{D_L def}
\end{align}
where $D(x,y)$ is the kernel of the Dirac operator in the infinite lattice $\mathbb{Z}^4$. The quantum theory is then
completely characterized by the expectational functional defined by the fermionic path-integration:
\begin{align}
\label{eq:Expect-Dirac}
\Expect{F}:=\frac{1}{Z}\int{\cal D}[\psi]{\cal D}[ \bar \psi] \, {\rm e}^{A(\bar\psi,\psi)} F(\bar\psi,\psi), 
\end{align}
where the Grassmann integration for each field variable is specified as
\begin{align}
\int d \psi_\alpha(x) \psi_\alpha(x) =1,\qquad \int d \bar\psi_\alpha(x) \bar\psi_\alpha(x) =1, 
\end{align}
and the functional measure is defined by 
\begin{align} 
\label{eq:Dirac-fermion-measure}
{\cal D}[\psi]{\cal D}[ \bar \psi]  :=
\prod_{x \in \Lambda ; \alpha=1,2,3,4} \{ d \psi_\alpha(x)  d \bar \psi_\alpha(x)  \}. 
\end{align}

The reflection positivity condition --- a condition on this expectational functuonal --- is formulated as follows: let us define time reflection operator $\theta$
which acts on polynomials of the fermionic field variables by the relations
\begin{align}
\theta(\psi(x))=\left( \bar\psi(\theta x)\gamma_0 \right)^T ,\quad
\theta(\bar\psi(x))=\left( \gamma_0\psi(\theta x) \right)^T \label{eq:t-reflection-psi}\\
\theta(\alpha F +\beta G)=\alpha^* \theta(F) +\beta^*\theta(G),\quad
\theta(FG)=\theta(G)\theta(F),
\end{align}
where we denote $\theta(t,\bvec x)=(-t+1,\bvec x)$ and $F,G$ are arbitrary polynomials of fermionic fields
and * means complex conjugation.
Let $\Lambda_\pm\subset\Lambda$ be the sets of sites with positive or non-positive time respectively.
Let $\mathcal{A_\pm}$ be the algebra of all the polynomials of the fields on $\Lambda_\pm$, and $\mathcal{A}$ 
on $\Lambda$. 
Then one says the theory is reflection positive if 
its expectation $\Expect\cdot:\mathcal{A}\to\mathbb{C}$ satisfies 
\begin{align}
\Expect{\theta(F_+)F_+}\ge0   \quad  \text{for} \, \,  {}^\forall F_+ \in \mathcal{A_+}  \label{RP}.
\end{align}

A popular choice of lattice Dirac operator is the Wilson Dirac operator, 
\begin{align}
D_{\rm w}
=
\sum_{\mu=0,1,2,3}\left\{ 
\frac{1}{2}\gamma_\mu(\partial_\mu-\partial_\mu^\dagger)+\frac{1}{2}\partial_\mu^\dagger\partial_\mu.
\right\},
\end{align}
%
Here we consider the overlap Dirac operator
\begin{align}
D&=\frac{1}{2}\Bigg(1+X\frac{1}{\sqrt{X^\dagger X}}\Bigg),\quad X=D_{\rm w}-m,
\end{align}
for $ 0 < m \le 1$.  This lattice Dirac operator  describes a single massless Dirac fermion and 
satisfies the GW relation, $\gamma_5 D + D \gamma_5 = 2 D \gamma_5 D$. Although 
the action is necessarily non ultra-local \cite{Horvath:1998cm}, 
the free overlap Dirac fermion indeed satisfies the reflection positivity condition, 
as will be shown below.\\


\noindent
\section{Proof of Reflection Positivity of overlap Dirac fermion}
%
%

To prove the reflection positivity, we need some additional 
definitions and notations.
First, let us denote 
\begin{align}
\Expect{F}_0:=  
 \int {\cal D}[\psi]{\cal D}[ \bar \psi] F(\bar\psi,\psi).
\end{align}
This $\Expect{\cdot}_0$ defines a linear function from ${\cal A}$ into $\mathbb{C}$.
Second, we decompose the lattice action $A$ into the following three parts :
\begin{align}
A=A_+ +A_- +\Delta A
\end{align}
where $A_+\in{\cal A_+}$, $A_-\in{\cal A_-}$, and $\Delta A$ is the part of the
action which contain both positive and negative time fields.
Thirdly, let us call $\mathcal{P}$ the set of all polynomials of the form 
$\sum_{j}\theta(F_{+ j})F_{+ j}$ in a finite summation, where $F_{+ j}\in\mathcal{A}_+$.

Although the above definition of ${\cal P}$ works well for the proof of the Wilson fermion, 
it is not enough for the proof of the overlap fermion. 
In our case of the overlap fermion, 
one needs to consider not only finite summations of the form $\sum_{j}\theta(F_{+ j})F_{+ j}$, 
but also infinite summations or integrations like 
\begin{align}\label{Riemanian sum}
\int ds\,\theta(F(s))F(s)=\lim_{N\to\infty}\sum_{k=1}^N \theta(F(s_k))F(s_k)\Delta s_k,
\end{align}
where the integration is defined as a limit of a finite Riemanian summation
 (see also eq. \eqref{Delta A positive}). 
%
To this end, we consider ${\cal \bar P}$, the closure of ${\cal P}$. The closure ${\cal \bar P}$
contains not only elements of the original ${\cal P}$, but also all the limit points of conversing sequences in ${\cal P}$.
That is, 
\begin{align}
F\in{\cal\bar P}  \quad \Leftrightarrow \quad  {}^\exists \{ F_n \}_{n=1}^\infty \in{\cal P} \, : \, \lim_{n\to\infty}F_n=F.
\end{align}
Here, the sequence $\{F_n\}_n \subset{\cal A}$ is defined to be convergent to some $F\in{\cal A}$, if 
any coefficient in $F_n$ converges to the corresponding coefficient in $F$ as a complex number
\footnote{
This definition of convergence in ${\cal A}$ is equivalent to the norm convergence of the Grassmann
algebra induced from the metric of the underlying vector space, where the fermionic fields
form an orthonomal basis.
}.
Note that with respect to this definition of convergence, 
the linear operation, the product operation
in $\mathcal{A}$, and the linear mappings $\Expect\cdot_0,\Expect\cdot:\mathcal{A}\to\mathbb{C}$ 
are all continuous functions, i.e.
if $F_n\to F$, $G_n\to G$, then 
\begin{align}
\alpha F_n +\beta G_n\to \alpha F +\beta G, \quad F_n G_n\to FG, \label{conti of prod and linear combi}\\
\Expect{F_n}\to\Expect{F},\quad \Expect{F_n}_0\to\Expect{F}\label{conti of expectations}.
\end{align}
%

Now, 
we note the fact  that the following four statements (i)-(iv) imply the reflection positivity: 

\vspace{.5em}
\noindent
(i) If $F,G$ belong to ${\cal\bar P}$ then $FG$ also belongs to ${\cal\bar P}$.  \\
(ii) For all $F\in{\cal\bar P}$, $\Expect{F}_0\ge0$.\\
(iii) $\theta(A_+)=A_-$.\\
(iv) $\Delta A\in{\cal \bar P}$.

\vspace{.5em}
\noindent
In fact, from these statements, it follows that 
\begin{align}\label{zero positivity}
\Expect {{\rm e}^A\,\theta(F_+)F_+ }_0 &=\Expect {{\rm e}^{A_+ +A_- +\Delta A}\,\theta(F_+)F_+ }_0 
=\Expect {{\rm e}^{A_+ +\theta(A_+) +\Delta A}\,\theta(F_+)F_+ }_0 \no\\
&=\Expect {\underbrace{\theta({\rm e}^{A_+}){\rm e}^{A_+}\,{\rm e}^{\Delta A}}_{\in\mathcal{\bar P}\;\text{(by (i),(iv))}}
\underbrace{\theta(F_+)F_+}_{\in\mathcal{\bar P}} }_0\ge 0  
\end{align}
 for arbitrary $F_+\in\mathcal{A_+}$. Considering the special case where $F_+=1\in\mathcal{A}_+$, we have
$\Expect{{\rm e}^A}_0\ge 0$. Hence, we obtain
\begin{align}\label{zero to general}
\Expect {\theta(F_+)F_+}&= \frac{\Expect {{\rm e}^A\,\theta(F_+)F_+ }_0}{\Expect{{\rm e}^A}_0}\ge 0.
\end{align}
Therefore 
the proof is reduced to showing these four statements (i)-(iv).

Next, we will give the proofs of the statements (i)-(iv).
The statement (i) follows from the similar statement with ${\cal P}$, 
which has been proved for the Wilson case \cite{Osterwalder:1977pc} . 

To show the statement (ii), one should refer to the definition of fermionic integration measure. 
With the definition (\ref{eq:Dirac-fermion-measure}),  it is sufficient to consider  
$F_+\in\mathcal{A}_+$ of the form
\begin{align}
F_+ = \prod_{x\in\Lambda_+ ; \alpha=1,2,3,4} \{ \bar \psi_\alpha(x)  \psi_\alpha(x)   \} \in \mathcal{P},  
\end{align}
for which one can see  
\begin{align}
\int {\cal D}[\psi]{\cal D}[ \bar \psi] \,  \theta(F_+) F_+ = \{ \det (\gamma_0^2) \}^{16L^4}= 1\ge 0.
\end{align}
Therefore, one concludes that for arbitrary $F\in\mathcal{P}$, $\Expect{F}_0\ge 0$.
Take arbitrary $F\in\mathcal{\bar P}$. Then there exists a converging sequence $\{F_n\}_n$ such that $F_n\to F$.
From the continuity of $\Expect{\cdot}_0$ (see \eqref{conti of expectations}), we obtain
\begin{align}
\Expect{F}_0=\Expect{\lim_{n\to\infty}F_n}_0=\lim_{n\to\infty}\Expect{F_n}_0\ge 0.
\end{align}

 The statement (iii) can be shown by using the property of 
the overlap Dirac kernel: $D_L^\dagger(x,y)=\gamma_0D_L(\theta x,\theta y)
\gamma_0$. 

To show the statement (iv) $\Delta A\in\bar{\mathcal{P}}$, 
we use a spectral representation of $D_L(x,y)$.
To derive the spectral representation of $D_L$, we first Fourier transform 
the  overlap Dirac operator kernel $D(x,y)$ in the infinite volume:
\begin{align}
\label{eq-FT of D}
D(x,y)\Big|_{x_0\not=y_0}=\int\frac{d^4\bvec p}{(2\pi)^4}\,{\rm e}^{ip\cdot(x-y)}\frac{X(p_0,\bvec p)}{2\sqrt{X^\dagger X(p_0,\bvec p)}},
\end{align}
where $X(p_0,\bvec p)=\sum_{\mu}i\gamma_\mu \sin p_\mu +\sum_\mu(1-\cos p_\mu)-m$.
Then, we change the $p_0$ integration region, $[-\pi,\pi]$, to the contours along the imaginary axis
in the complex $p_0$ plane by Cauchy's integration theorem, as shown in 
FIG.~\ref{fig:p0-integration-contour}. 
Depending whether $x_0 - y_0 > 0$ or  $x_0 - y_0 < 0$, 
we choose the contours 
$[i E_1, i \infty]$ or  $[-i E_1,- i \infty]$, respectively, to obtain
\begin{align}
D(x,y)\Big|_{x_0-y_0>0}
&=\int\frac{d^3\bvec p}{(2\pi)^3}\int_{E_1}^\infty\frac{dE}{2\pi}\,{\rm e}^{-E(x_0-y_0)} {\rm e}^{i\bvec p\cdot (\bvec x-\bvec y)}
\frac{X(iE,\bvec p)}{\sqrt{-X^\dagger X(iE,\bvec p)}}\label{D positive}\\
D(x,y)\Big|_{x_0-y_0<0}
&\quad=\int\frac{d^3\bvec p}{(2\pi)^3}\int_{E_1}^\infty\frac{dE}{2\pi}\,{\rm e}^{E(x_0-y_0)} {\rm e}^{i\bvec p\cdot (\bvec x-\bvec y)}
\frac{X(-iE,\bvec p)}{\sqrt{-X^\dagger X(iE,\bvec p)}}.\label{D negative}
\end{align}
where $E_1$ is the edge of the cut coming from the square root, and is determined by the relations
\begin{align}
X^\dagger X(iE_1,\bvec p)=0, \quad E_1>0.
\end{align}
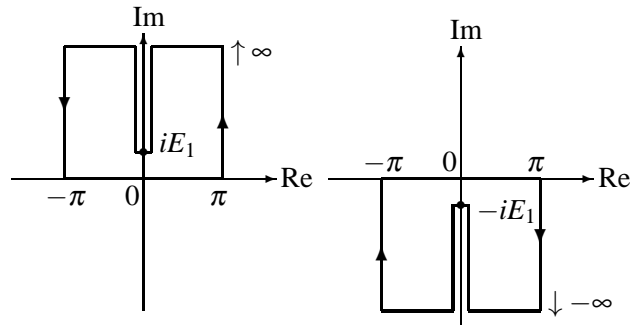
\begin{figure}[b]
\begin{center}
\begin{picture}(250,120)
\put(0,60){\vector(1,0){100}}
\put(50,10){\vector(0,1){105}}
\thicklines
\put(20,60){\line(1,0){60}}
\thicklines
\put(20,60){\line(1,0){60}}
\thicklines
\put(80,60){\vector(0,1){25}}
\thicklines
\put(80,85){\line(0,1){25}}
\thicklines
\put(53,110){\line(1,0){27}}
\thicklines
\put(53,110){\line(0,-1){40}}
\thicklines
\put(53,70){\line(-1,0){6}}
\thicklines
\put(47,70){\line(0,1){40}}
\thicklines
\put(47,110){\line(-1,0){27}}
\thicklines
\put(20,110){\vector(0,-1){25}}
\thicklines
\put(20,85){\line(0,-1){25}}

\put(43,50){0}
\put(75,50){$\pi$}
\put(13,50){$-\pi$}
\put(56,70){$iE_1$}
\put(50,70){\circle*{3}}
\put(102,57){Re}
\put(46,118){Im}
\put(83,105){$\uparrow\infty$}

\thinlines
\put(120,60){\vector(1,0){100}}
\put(170,5){\vector(0,1){105}}
\thicklines
\put(140,60){\line(1,0){60}}
\thicklines
\put(140,60){\line(1,0){60}}
\thicklines
\put(200,60){\vector(0,-1){25}}
\thicklines
\put(200,35){\line(0,-1){25}}
\thicklines
\put(173,10){\line(1,0){27}}
\thicklines
\put(173,10){\line(0,1){40}}
\thicklines
\put(173,50){\line(-1,0){6}}
\thicklines
\put(167,50){\line(0,-1){40}}
\thicklines
\put(167,10){\line(-1,0){27}}
\thicklines
\put(140,10){\vector(0,1){25}}
\thicklines
\put(140,35){\line(0,1){25}}

\put(163,63){0}
\put(195,63){$\pi$}
\put(133,63){$-\pi$}
\put(176,45){$-iE_1$}
\put(170,50){\circle*{3}}
\put(222,57){Re}
\put(166,113){Im}
\put(203,10){$\downarrow-\infty$}
\end{picture}
\caption{Complex integration contours}
\label{fig:p0-integration-contour}
\end{center}
\end{figure}

In this spectrum representaion of $D$, 
it is very crucial to notice the fact that $\mp \gamma_0 X(\pm iE,\bvec p)$ ($E\ge E_1$) are positive definite matrices  
and there exist matrices $Y_{\pm}(E,\bvec p)$ such that
\begin{align}
\label{gamma0-X-positive}
\mp \gamma_0 X(\pm iE,\bvec p)=Y_{\pm}^\dagger Y_{\pm}(E,\bvec p) \quad (E\ge E_1). 
\end{align}
In fact, it is not difficult to check that $Y_{\pm}(E,\bvec p)$ are given by
\begin{align}
Y_{\pm}(E,\bvec p)=-\sum_{k=1}^3\frac{l(E,\bvec p)\sin p_k}{W(E,\bvec p)}\gamma_k \mp i\frac{W(E,\bvec p)}{2l(E,\bvec p)}\gamma_0+il(E,\bvec p), \no
\end{align}
where $W(E,\bvec p)=\sum_{k=1}^3(1-\cos p_k) +1 -\cosh E -m$ and 
\begin{align}
l(E,\bvec p)&= \left[ \frac{1}{2}\frac{\sinh E}{\sum_{k=1}^3\sin^2 p_k/W(E,\bvec p)^2+1}
\Bigg(1+ \sqrt{1-\frac{\sum_{k=1}^3\sin^2 p_k+W(E,\bvec p)^2}{\sinh^2 E}}\Bigg)\right]^{\frac{1}{2}}. 
\no
\end{align} 

From the equations \eqref{D_L def}, \eqref{D positive} and \eqref{D negative}, 
we find the spectrum representation of $D_L(x,y)$ as follows: putting $V=1/(2L)^3$, 
\begin{align}
D_L(x,y)\Big|_{x_0\not=y_0}&=\sum_{\bvec p}\int_{E_1}^\infty\frac{dE}{2\pi}\frac{1}{1+{\rm e}^{-2EL}}\frac{1}{V}
 {\rm e}^{-E|x_0-y_0|}{\rm e}^{i\bvec p\cdot (\bvec x-\bvec y)}
\frac{X(\epsilon iE,\bvec p)}{\sqrt{-X^\dagger X(iE,\bvec p)}}\no\\
&\qquad+\sum_{\bvec p}\int_{E_1}^\infty\frac{dE}{2\pi}\frac{{\rm e}^{-2EL}}{1+{\rm e}^{-2EL}}\frac{1}{V}
{\rm e}^{E|x_0-y_0|}{\rm e}^{i\bvec p\cdot (\bvec x-\bvec y)}
\frac{-X(-\epsilon iE,\bvec p)}{\sqrt{-X^\dagger X(iE,\bvec p)}},\label{D_L}
\end{align}
where $\epsilon$ is defined as the sign of $x_0-y_0$, and the spacial momentum $p_k$ runs over $p_k=n_k\pi/L,\,(-L\le n\le L)$
in the above summation. 
In \eqref{D_L}, 
the first term  becomes $D(x,y)$ in the limit $L\to\infty$, and
the second term represents a `finite lattice effect' which vanishes in the limit $L\to\infty$. 
The latter is the contribution of the  wrong-sign-energy modes and 
the minus sign appearing in front of $X(-\epsilon iE,\bvec p)$
comes from the \textit{anti}-periodicity in the time direction, which  
is required 
for the positivity,  as will be seen. 



From these observations, now we can show that $\Delta A\in\bar{\mathcal{P}}$: 
for the term with $x_0>0$, $y_0\le 0$ (in this case $\epsilon=1$), we obtain
\begin{align}
\sum_{x\in\Lambda_+}\sum_{y\in\Lambda_-}\bar\psi(x)D_L(x,y)\psi(y)
&=-\sum_{\bvec p}\int_{E_1}^\infty\frac{dE}{2\pi}\frac{1}{V}\Bigg[C_{E,\bvec p}\theta(C_{E,\bvec p})+D_{E,\bvec p}\theta(D_{E,\bvec p})\Bigg],\label{Delta A positive}
\end{align}
where $C_{E,\bvec p}$ and $D_{E,\bvec p}$ are defined by
\begin{align}
C_{E,\bvec p}
&=\sqrt{\frac{1}{1+{\rm e}^{-2EL}}}\sum_{x\in\Lambda_+}\bar\psi(x)\gamma_0\tilde{Y_+}(E,\bvec p)^\dagger 
{\rm e}^{-Ex_0}{\rm e}^{i\bvec p\cdot\bvec x},\\
D_{E,\bvec p}
&=\sqrt{\frac{{\rm e}^{-2EL}}{1+{\rm e}^{-2EL}}}\sum_{x\in\Lambda_+}\bar\psi(x)\gamma_0\tilde{Y_-}(E,\bvec p)^\dagger 
{\rm e}^{Ex_0}{\rm e}^{i\bvec p\cdot\bvec x},
\end{align}
with
$\tilde{Y_+}(E,\bvec p)=Y_+(E,\bvec p)/(-X^\dagger X(iE,\bvec p))^{\frac{1}{4}}$.
The overall minus sign in the r.h.s. of \eqref{Delta A positive} results 
from \eqref{D_L} by using \eqref{gamma0-X-positive}. 
This minus sign is canceled after exchanging
the order of the Grassmann products 
in \eqref{Delta A positive}, 
and we see that this term belongs to $\bar{\mathcal{P}}$.
%
One can show similarly that the term with $x_0\le0$, $y_0>0$ (in this case $\epsilon=-1$) also belongs to $\bar{\mathcal{P}}$. 
Thus we obtain $\Delta A\in\bar{\mathcal{P}}$ and complete the proof of the reflection positivity.

\section*{Acknowledgements}
K.U. would like to thank Tsutomu T. Yanagida for valuable encouragements.
K.U. is supported by Global COE Program ``the Physical Science Frontier", MEXT, Japan.
This work was supported by World Premier International Center Initiative (WPI Program), MEXT, Japan.
Y.K. would like to thank S.~Hashimoto and H.~Fukaya for discussions.
Y.K. is supported in part by Grant-in-Aid for Scientific Research No.~21540258, ~21105503.

\end{document}